\documentclass[12pt,a4paper]{article}
\usepackage[british]{babel}
\usepackage{epsfig}
                                                                                
\begin{document}                                                                
\date{}

\title{
{\vspace{-20mm} \normalsize
\hfill \parbox[t]{50mm}{DESY 03-020}}\\[20mm]
{\Large\bf A recurrence scheme for least-square \\
           optimized polynomials}               \\[2mm]}
\author{ C. Gebert and I. Montvay                        \\
         Deutsches Elektronen-Synchrotron DESY           \\
         Notkestr.\,85, D-22603 Hamburg, Germany }

\newcommand{\be}{\begin{equation}}                                              
\newcommand{\ee}{\end{equation}}                                                
\newcommand{\half}{\frac{1}{2}}                                                 
\newcommand{\rar}{\rightarrow}                                                  
\newcommand{\lar}{\leftarrow}
\newcommand{\LCB}{\raisebox{-0.3ex}{\mbox{\LARGE$\left\{\right.$}}}
\newcommand{\RCB}{\raisebox{-0.3ex}{\mbox{\LARGE$\left.\right\}$}}}
\newcommand{\U}{\mathrm{U}}
\newcommand{\SU}{\mathrm{SU}}
                                                                                
\maketitle
\vspace*{1em}

\begin{abstract} \normalsize
 A recurrence scheme is defined for the numerical determination of
 high degree polynomial approximations to functions as, for instance,
 inverse powers near zero.
 As an example, polynomials needed in the two-step multi-boson (TSMB)
 algorithm for fermion simulations are considered.
 For the polynomials needed in TSMB a code in C is provided which is
 easily applicable to polynomial degrees of several thousands.
\end{abstract}       

\section{Introduction}\label{sec1}

 High quality polynomial approximations of inverse powers near zero are
 needed, for instance, in {\em multi-bosonic algorithms} for numerical
 Monte Carlo simulations of fermionic quantum field theories \cite{MB}.
 Least-square optimization offers an effective and flexible framework,
 in particular in {\em two-step multi-boson} algorithms \cite{TSMB},
 where the functions to be approximated are typically of the form
 $x^{-\alpha}\bar{P}(x)$ with $\alpha > 0$ and some given polynomial
 $\bar{P}(x)$.
 The approximation has to be optimized in an interval covering the
 spectrum of a positive matrix, that is for
 $x \in [\epsilon,\lambda],\; 0 \leq \epsilon < \lambda$.
 The polynomial $\bar{P}(x)$ can be just a constant or another cruder
 approximation of $x^{-\alpha}$.

 The required precision of the approximation is rather high, it should
 be preferably near machine precision.
 Since the {\em condition number} $\lambda/\epsilon$ can be as high as
 $10^6$ or more, the required polynomial degrees can be in the multi
 thousand range.
 This is due to the fact that -- with
 $\lambda = {\cal O}(1)-{\cal O}(10)$ -- the lower limit is of the order
 $\epsilon = (am_f)^2$ where $am_f$ is the mass of the elementary
 fermion in lattice units.
 Interesting values for the fermion mass are, for instance in QCD or in
 supersymmetric Yang-Mills theory, $am_f = {\cal O}(10^{-3})$ or smaller.
 (For examples of recent TSMB applications with light fermions see
 \cite{SYMWTI} and \cite{LIGHTQ,VALENCE}.)
 Other possible applications of high degree least-square optimized
 polynomial approximations are, for instance, in fermionic simulation
 algorithms for improved actions \cite{KNECHTHASEN,ALEXHASEN} or the
 evaluation of the matrix elements of Neuberger's overlap fermion action
 \cite{NEUBERGER}.

 Least-square optimized polynomials in the context of the TSMB algorithm
 have been considered previously in \cite{POLYNOM,WUPPER}.
 (For a general introduction see, for instance, \cite{APPROXIMA}.)
 In the present paper the recurrence scheme proposed in \cite{WUPPER}
 is worked out in detail and an optimized C code is provided which is
 suitable to deliver the required polynomials with an acceptable amount
 of computer time.

 The plan of this paper is as follows: in the next section some basic
 definitions are briefly summarized and the notations are introduced.
 In section \ref{sec3} the recurrence scheme for the numerical
 determination of the least-square optimized polynomials is presented in
 detail.
 Section \ref{sec4} is devoted to the explanation of the C code which
 can be used to perform the recurrence.

\section{Definitions}\label{sec2}

 In this section the basic definitions and notations are introduced
 which will be needed later.
 For more formulas and discussion see \cite{POLYNOM} and especially
 \cite{WUPPER} where a rather complete collection of formulas is given
 (and a few misprints in \cite{POLYNOM} are also corrected).

 In the general case one wants to approximate the real function $f(x)$
 in the interval $x \in [\epsilon,\lambda]$ by a polynomial $P_n(x)$ of
 degree $n$.
 The aim is to minimize the deviation norm
\begin{equation} \label{eq01}
\delta_n \equiv \left\{ N_{\epsilon,\lambda}^{-1}\;
\int_\epsilon^\lambda dx\, w(x)^2 \left[ f(x) - P_n(x) \right]^2
\right\}^\half \ .
\end{equation}
 Here $w(x)$ is an arbitrary real weight function and the overall
 normalization factor $N_{\epsilon,\lambda}$ can be chosen by
 convenience, for instance, as
\begin{equation} \label{eq02}
N_{\epsilon,\lambda} \equiv
\int_\epsilon^\lambda dx\, w(x)^2 f(x)^2 \ .
\end{equation}
 For optimizing the relative deviation one takes a weight function
 $w(x) = f(x)^{-1}$.

 As discussed in detail in ref.~\cite{WUPPER}, obtaining the optimized
 polynomial $P_n(x)$ in case of $x \in [\epsilon,\lambda]$ with small
 positive $\epsilon$ and $f(x)=x^{-\alpha}/\bar{P}(x)$ ($\alpha > 0$
 and polynomial $\bar{P}(x)$) is equivalent to the inversion of an 
 ill-conditioned matrix with a very large condition number.
 As a consequence, the problem can only be solved by using high
 precision arithmetics.

 Another place where rounding errors may be disturbing is the evaluation
 of the polynomials -- once the polynomial coefficients of $P_n(x)$ are
 known.
 Fortunately, this second problem can be solved without the need of high
 precision arithmetics by applying a suitable orthogonal expansion.
 In fact, after the optimized polynomials are obtained with high
 precision arithmetics, their evaluation can be performed even with
 32-bit arithmetics if this orthogonal expansion is used.
 The orthogonal polynomials $\Phi_\mu(x)\; (\mu=0,1,2,\ldots)$
 are defined in such a way that they satisfy
\begin{equation} \label{eq03}
\int_\epsilon^\lambda dx\, w(x)^2 \Phi_\mu(x)\Phi_\nu(x)
= \delta_{\mu\nu} q_\nu \ .
\end{equation}
 The polynomial $P_n(x)$ can be expanded in terms of them according to
\begin{equation} \label{eq04}
P_n(x) = \sum_{\nu=0}^n d_{n\nu} \Phi_\nu(x) \ .
\end{equation}
 Besides the normalization factor $q_\nu$ let us also introduce, for
 later purposes, the integrals $p_\nu$ and $s_\nu$ ($\nu=0,1,\ldots$) by
\begin{eqnarray} \label{eq05}
q_\nu &\equiv& \int_\epsilon^\lambda dx\, w(x)^2 \Phi_\nu(x)^2 \ ,
\nonumber \\[0.5em]
p_\nu &\equiv& \int_\epsilon^\lambda dx\, w(x)^2 \Phi_\nu(x)^2 x \ ,
\nonumber \\[0.5em]
s_\nu &\equiv& \int_\epsilon^\lambda dx\, w(x)^2 x^\nu \ .
\end{eqnarray}

 It can be easily shown that the expansion coefficients $d_{n\nu}$
 minimizing $\delta_n$ are independent of $n$ and are given by
\begin{equation} \label{eq06}
d_{n\nu} \equiv d_\nu = \frac{b_\nu}{q_\nu} \ ,
\end{equation}
 where
\begin{equation} \label{eq07}
b_\nu \equiv \int_\epsilon^\lambda dx\, w(x)^2 f(x) \Phi_\nu(x) \ .
\end{equation}
 The minimum of the deviation norm with normalization defined in
 (\ref{eq02}) is:
\begin{equation} \label{eq08}
\delta_n^{min} = \left\{ 1 -  N_{\epsilon,\lambda}^{-1}\;
\sum_{\nu=0}^n \frac{b_\nu^2}{q_\nu} \right\}^\half \ .
\end{equation}

 The orthogonal polynomials satisfy three-term recurrence
 relations \cite{APPROXIMA}.
 The first two of them with $\mu=0,1$ are given by
\begin{equation} \label{eq09}
\Phi_0(x) = 1 \ , \hspace{2em}
\Phi_1(x) = x - \frac{s_1}{s_0} \ .
\end{equation}
 The higher order polynomials $\Phi_\mu(x)$ for $\mu=2,3,\ldots$ can be
 obtained from the recurrence relation
\begin{equation} \label{eq10}
\Phi_{\mu+1}(x) = (x+\beta_\mu)\Phi_\mu(x) +
\gamma_{\mu-1}\Phi_{\mu-1}(x) \ ,
\hspace{3em} (\mu=1,2,\ldots) \ ,
\end{equation}
 where the recurrence coefficients are given by
\begin{equation} \label{eq11}
\beta_\mu = -\frac{p_\mu}{q_\mu} \ , \hspace{3em}
\gamma_{\mu-1} = -\frac{q_\mu}{q_{\mu-1}} \ .
\end{equation}
%

\section{Recurrence scheme}\label{sec3}

 A recurrence scheme is defined as a sequential algorithm for building
 up the system of orthogonal polynomials and determining the expansion
 coefficients of the optimized polynomial approximation starting
 from the function $f(x)$ to be approximated and from the weight
 function $w(x)$.
 The necessary ingredients are, in fact, the moments of the weight
 function $s_\nu$ defined in (\ref{eq05}) and the moments of
 $w(x)^2 f(x)$:
\begin{equation} \label{eq12}
t_\nu \equiv \int_\epsilon^\lambda dx\, w(x)^2 f(x) x^\nu \ ,
\hspace{3em}
(\nu=0,1,\ldots ) \ .
\end{equation}
 Since the calculations have to be performed in high precision
 arithmetics, these moments are needed to high precision which is easily
 reached if the integrals can be analytically performed.
 Otherwise the moments themselves have to be numerically calculated with
 high precision arithmetics.

 In case of the TSMB applications we typically have to consider the
 function $f(x) = x^{-\alpha}/\bar{P}(x)$.
 In this case, if one chooses to optimize the relative deviation,
 the basic integrals defined in (\ref{eq05}) and (\ref{eq12}) are,
 respectively,
\begin{eqnarray} \label{eq13}
s_\nu & = & \int_\epsilon^\lambda dx\, \bar{P}(x)^2 x^{2\alpha+\nu} \ ,
\nonumber \\[0.5em]
t_\nu & = & \int_\epsilon^\lambda dx\, \bar{P}(x) x^{\alpha+\nu} \ .
\end{eqnarray}
 These integrals can be simplified if, instead of choosing the weight
 factor for relative deviation $w(x)^2=\bar{P}(x)^2 x^{2\alpha}$,
 one takes its square root:
\begin{equation} \label{eq14}
w(x)^2 = \bar{P}(x) x^\alpha \ ,
\end{equation}
 In practice this leads to similarly good approximations as the previous
 choice and the basic integrals become simpler:
\begin{eqnarray} \label{eq15}
s_\nu & = & \int_\epsilon^\lambda dx\, \bar{P}(x) x^{\alpha+\nu} \ ,
\nonumber \\[0.5em]
t_\nu & = & \int_\epsilon^\lambda dx\, x^\nu \ .
\end{eqnarray}

 One can devise recurrence schemes differently, for instance, based on
 the polynomial coefficients of the orthogonal polynomials.
 A good scheme can be built on the moments
\begin{eqnarray} \label{eq16}
& & r_{\mu\nu} \equiv
\int_\epsilon^\lambda dx\, w(x)^2 \Phi_\mu(x) x^\nu \ ,
\nonumber \\[0.5em]
& & \mu=0,1,\ldots,n;
\hspace{1.5em} \nu=\mu,\mu+1,\ldots,2n-\mu \ ,
\end{eqnarray}
 using also the coefficients of the next-to-leading order term of the
 orthogonal polynomials.
 According to the recurrence relation in (\ref{eq10}) we have for
 $\mu=1,2,\ldots$
\begin{equation} \label{eq17}
\Phi_\mu(x) = x^\mu + f_\mu x^{\mu-1} + \cdots
\end{equation}
 where the next-to-leading order coefficient $f_\mu$ satisfies
\begin{equation} \label{eq18}
f_1 = -\frac{s_1}{s_0} \ , 
\hspace{3em}
f_{\mu+1} = f_\mu + \beta_\mu
\hspace{3em}
(\mu = 1,2,\ldots) \ .
\end{equation}

 The recurrence coefficients $\beta_\mu,\;\gamma_{\mu-1}$ can be
 expressed from
\begin{equation} \label{eq19}
q_\mu = r_{\mu\mu} \ ,\hspace{2em} 
p_\mu=r_{\mu,\mu+1} + f_\mu r_{\mu\mu}
\end{equation}
 and eq.~(\ref{eq11}) as
\begin{equation} \label{eq20}
\beta_\mu = - f_\mu - \frac{r_{\mu,\mu+1}}{r_{\mu\mu}} \ ,
\hspace{2em}
\gamma_{\mu-1} = - \frac{r_{\mu\mu}}{r_{\mu-1,\mu-1}} \ .
\end{equation}
 It follows from the definition (\ref{eq16}) that
\begin{eqnarray} \label{eq21}
r_{0\nu} & = & \int_\epsilon^\lambda dx\, w(x)^2 x^\nu = s_\nu \ ,
\nonumber \\[0.5em]
r_{1\nu} & = & \int_\epsilon^\lambda dx\, w(x)^2 
(x^{\nu+1} + f_1 x^\nu) = s_{\nu+1} + f_1 s_\nu 
= s_{\nu+1} - \frac{s_1 s_\nu}{s_0} \ .
\end{eqnarray}
 The recurrence relation (\ref{eq10}) for the orthogonal polynomials
 implies
\begin{equation} \label{eq22}
r_{\mu+1,\nu} = r_{\mu,\nu+1} + \beta_\mu r_{\mu\nu}
+ \gamma_{\mu-1}r_{\mu-1,\nu} \ .
\end{equation}
 This completes a closed recurrence scheme for calculating 
 $\beta_\mu$ and $\gamma_{\mu-1}$ which can then be used to determine
 the orthogonal polynomials by (\ref{eq10}).

 In order to see how this recurrence scheme works let us display a
 few steps explicitly.
 Let us assume that we want to obtain the orthogonal polynomials up to
 degree $n$ and we know already the basic integrals $s_\nu$ for
 $\nu=0,1,\ldots,2n$.
 According to (\ref{eq21}) we then have
\begin{equation} \label{eq23}
r_{0\nu} = s_\nu \hspace{1em} (\nu=0,1,\ldots,2n-1) \ ,\hspace{2em}
r_{1\nu} = s_{\nu+1} - \frac{s_1 s_\nu}{s_0} \hspace{1em}
(\nu=1,2,\ldots,2n-1) \ .
\end{equation}
 Using (\ref{eq18}) and (\ref{eq20}) we obtain
\begin{equation} \label{eq24}
f_1 = -\frac{s_1}{s_0} \ ,\hspace{2em}
\beta_1 = -f_1-\frac{r_{12}}{r_{11}} \ ,\hspace{2em}
\gamma_0 = -\frac{r_{11}}{r_{00}} \ .
\end{equation}
 Now (\ref{eq22}) gives
\begin{equation} \label{eq25}
r_{2\nu} = r_{1,\nu+1} + \beta_1 r_{1\nu} + \gamma_0 r_{1\nu} 
\hspace{2em} (\nu=2,3,\ldots,2n-2) \ .
\end{equation}
 This can be used for calculating
\begin{equation} \label{eq26}
f_2 = \beta_1 + f_1 \ ,\hspace{2em}
\beta_2 = -f_2-\frac{r_{23}}{r_{22}} \ ,\hspace{2em}
\gamma_1 = -\frac{r_{22}}{r_{11}} \ ,
\end{equation}
 and so on.
 At the end there is
\begin{equation} \label{eq27}
r_{n-1,\nu} = r_{n-2,\nu+1} + \beta_{n-2} r_{n-2,\nu} + 
\gamma_{n-3} r_{n-3,\nu} 
\hspace{2em} (\nu=n-1,n,n+1)
\end{equation}
 which gives
\begin{equation} \label{eq28}
f_{n-1} = \beta_{n-2} + f_{n-2} \ ,\hspace{2em}
\beta_{n-1} = -f_{n-1}-\frac{r_{n-1,n}}{r_{n-1,n-1}} \ ,\hspace{2em}
\gamma_{n-2} = -\frac{r_{n-1,n-1}}{r_{n-2,n-2}} \ .
\end{equation}
 Finally one calculates for $q_n$
\begin{equation} \label{eq29}
r_{nn} = r_{n-1,n+1} + \beta_{n-1} r_{n-1,n} + 
\gamma_{n-2} r_{n-2,n} \ . 
\end{equation}

 After determining the orthogonal polynomials one can also perform a
 similar recurrence for the expansion coefficients of the least-square
 optimized polynomials given in (\ref{eq06}).
 Now one assumes the knowledge of the moments $t_\nu$ in (\ref{eq12})
 and considers
\begin{eqnarray} \label{eq30}
& & b_{\mu\nu} \equiv
\int_\epsilon^\lambda dx\, w(x)^2 f(x) \Phi_\mu(x) x^\nu \ ,
\nonumber \\[0.5em]
& & \mu=0,1,\ldots,n;
\hspace{1.5em} \nu=0,1,\ldots,n-\mu \ .
\end{eqnarray}
 From this one obtains $b_\nu = b_{\nu 0}$.
 For starting the iteration one has
\begin{equation} \label{eq31}
b_{0\nu} = t_\nu \ ,\hspace{2em}
b_{1\nu} = t_{\nu+1} + f_1 t_\nu \ ,
\end{equation}
 and then one uses
\begin{equation} \label{eq32}
b_{\mu+1,\nu} = b_{\mu,\nu+1} + \beta_\mu b_{\mu\nu}
+ \gamma_{\mu-1}b_{\mu-1,\nu} \ .
\end{equation}

 The basic recurrence relation in (\ref{eq10}) can also be used for the
 evaluation of integrals of least-square optimized polynomials.
 Examples are $t_\nu$ in (\ref{eq13}) and $s_\nu$ in (\ref{eq14}) if
 $\bar{P}(x)$ is expanded in the corresponding orthogonal basis.

\section{C code for polynomials needed in TSMB}\label{sec4}

 The implementation of the presented recurrence scheme has to be done with
 a sufficiently high numerical precision. We have found that the required 
 number of digits increases linearly with the polynomial order parameter.
 Since this polynomial degree can be in the multi thousand range one
 has to carefully choose a way to implement this. In particular it is
 obvious that the standard 32-bit or 64-bit precision, corresponding to roughly
 ${\cal O}(10)$ digits of precision, is insufficient for this task.

 Possible environments are computer algebra systems like Maple or
 more low-level computer programming languages like C or Fortran combined
 with a suitable library for multiprecision arithmetics. Since runtime
 performance is important we have opted for the programming language C.
\begin{figure*}
\begin{center}
\includegraphics[angle=-90,width=0.75\textwidth]{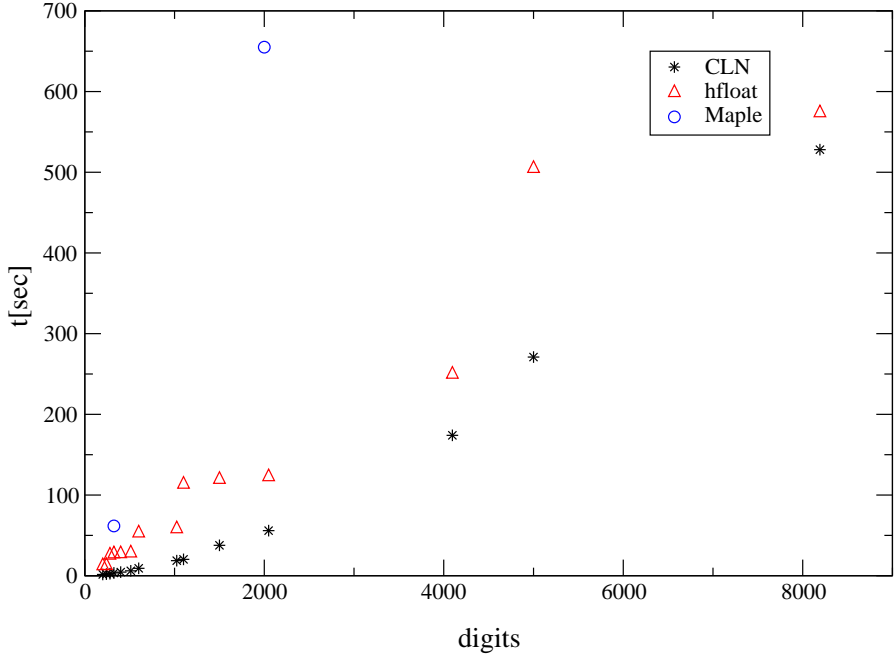}
\vspace*{-1em}
\parbox{12cm}{\caption{\label{fig_1}
 Comparison of the runtime performance when calculating the first three
 polynomials $P^{(1)}$, $P^{(2)}$ and $P^{(3)}$.}}
\end{center}
\vspace*{1.5em}
\end{figure*}

 There are several freely available libraries for multiprecision
 arithmetics \cite{CLN,GMP,HFLOAT}.
 These libraries vary in usability and performance.
 For very large numbers, i.~e.\ more than 12000 decimal digits, these
 libraries use the Sch\"onhage-Strassen multiplication, an
 asymptotically optimal algorithm.
 However, most of the time we need polynomial orders of
 ${\cal O}(1000)$, for which less than 5000 digits are needed.
 Therefore other algorithms like the ${\cal O}(n^2)$ standard algorithm
 for smaller numbers or the ${\cal O}(n^{1.6})$ Karatsuba multiplication
 for intermediate large numbers should be used where applicable.
 This is done by the gmp and the CLN libraries, where the latter comes
 with easy to use functions for the calculation of sine, cosine and
 logarithm.
 Furthermore it can be configured to use the low-level routines of gmp,
 so it can be considered as a nice interface for that library.
 A comparison of different environments and libraries is shown in
 figure~\ref{fig_1}.
 The polynomial degree was fixed for this performance test, while only
 the number of used digits was changed.
 Runtime measurements are for the computer algebra system Maple, and for
 the programming language C using the libraries hfloat and CLN.
 In this example the library CLN was used without support for the
 faster low-level routines of gmp.
 When interpreting the results of this comparison it has to be kept in
 mind that no guarantee for a similar level of optimization for the
 different programs can be given (although it should be roughly true). 
 Let us also note that in Maple one could switch the precision to lower
 values when sufficient, while this would not be so easy for the C
 libraries.
 This would allow to gain in performance for the Maple program but it
 would make a comparison with the other programs complicated.

 The source code for the program that uses the CLN library is 
 {\em quadropt.C}.
 It can be compiled with the {\em g++} compiler\footnote{
 To be precise, the program indeed has to be compiled with a C++
 compiler, because it uses C++ features like streams.
 The structure of the program is, however, very C-like, i.~e.\ it is
 based on usual function calls instead of calls to an objects member
 function.}
 and it has to be linked to the previously installed library CLN.
 The executable needs one parameter as an argument, a filename.
 This file is usually {\em quadropt.dat}, from which the options like
 polynomial degrees and interval are read.
 These options will be explained in the next subsections.
 The parameter file {\em quadropt.dat} is organized to have a line of
 comments and then a line of corresponding arguments.
 This structure is repeated for all arguments (see figure \ref{fig_2}).
\begin{figure*}
\begin{center}
{\small
\begin{verbatim}
Polynomial Order parameters(5+):
  16   60  10  90  20 30 70 90
Alpha, Epsilon, Lambda
 1.0   8.0e-3    4.0
W3(nom., denom.), MaxNeubergerTerms (<=0 -> P12-method), #P5-Newton iterations
      2  3      -1   1
Desired Precision (-1=default, +X=default+X), RootsPrecision (-1=default, etc.)
 -1  -1
Used data-subdirectory
 Data/Qcd2/Cn500.16.60
\end{verbatim}}
\vspace*{1em}
\parbox{12cm}{\caption{\label{fig_2}
 Structure of the parameter file {\em quadropt.dat}.}}
\end{center}
\vspace*{0.5em}
\end{figure*}

\subsection{The polynomial $P^{(1)}$}\label{sec4.1}

 The polynomial $P^{(1)}_{n_1}$ is supposed to be a polynomial
 approximation to the function $f(x)=x^{-\alpha}$, i.~e.\
 $\bar P(x)\equiv 1$. As weight function $\omega(x)^2$ the squared inverse
 of $f(x)$ is taken.
 The parameter $\alpha$ is given by the first value in the second line
 of arguments (i.~e.\ the fourth line altogether) of the parameter file.
 In case of the TSMB applications this is $\alpha=\frac{N_f}{2}$, where
 $N_f$ is originally the number of fermionic flavours.
 However, $N_f$ does not have to be integer valued, for instance, in
 the case of Majorana fermions or if determinant breakup is applied.
 In general any real value of $\alpha$ will work with the code.
 However, the necessary number of digits for a precise determination of
 the recurrence coefficients might differ if $\alpha$ is drastically
 different from those in the applications of the TSMB algorithm.

 In the first line of arguments the requested degrees for the different
 polynomials have to be given.
 The first value $n_1$ is for the polynomial degree of $P^{(1)}$.
 Although at least five numbers have to be specified it is possible to
 set the other values to zero.
 In this way only the first polynomial is calculated.

 Another important input is the approximation interval
 $[\epsilon,\lambda]$.
 The borders of this interval have to be given in the second line of
 arguments after $\alpha$.

 We have found that the required number of digits for the calculation of
 the necessary integrals and coefficients does only depend on the
 polynomial order.
 For the first polynomial $P^{(1)}$ it is sufficient to set the number
 of digits to $(40+1.6 n_1)$.
 One can further benefit by recognizing that some intermediate
 integrals need to be calculated with a lower precision of roughly
 $(30+0.9 n_1)$.
 Furthermore, the roots can be calculated with an even lower precison of
 roughly $(30+0.5 n_1)$.
 The situation might be different if the parameters $\alpha$, $\epsilon$
 and $\lambda$ are drastically different from those used in TSMB
 applications.

 In the C program as presented here the value of the number of digits is
 chosen according to the above rules.
 We do not exploit the possibility to use less digits at places where
 it would be in principle possible.
 The program might even use more digits if this is required by another
 calculated polynomial.
 Usually the necessary number of digits is determined by the fourth or
 fifth polynomial order.
 This is a drawback of the implementation in C where changing the number
 of digits is not really supported by the CLN library.
 In Maple one can set the number of digits as appropriate for each
 integral.
 Our C program uses a single fixed precision for all computations.
 The only exception is the ordering of the polynomial roots
 according to ref.~\cite{POLYNOM}.
 For this the calculation of a logarithm is needed roughly
 ${\cal O}(n^3)$ times.
 Since the computation of logarithms scales much worse than a simple
 multiplication, the ordering of the first polynomial roots could
 consume more than $90\%$ of the runtime already for modest values of
 $n_1$.
 Therefore the ordering is calculated by an appropriate lower precision,
 which is sufficient for this purpose.

 The default choice of the number of digits can be overruled
 by changing the two values in the fourth line of arguments from their
 default values of $-1$.
 With the first value the number of digits for the integrals can be
 specified.
 The second value gives the number of digits for the ordering of roots.
 If some value is set, then this is taken as the number of digits.
 If, however, this number starts with a $+$ character, e.~g.\ as in
 $+10$, then that number is added to the default number of digits.
 While $-1$ leads to the default number of digits, values smaller than
 $-2$ reduce the number of digits from their default value.
 A value of $-2$ as the second value sets the number of digits for the
 roots equal to the number of digits for the other integrals (this is
 highly not recommended).

 Of course one can compute the polynomials at a much higher precision
 than needed, e.~g.\ by giving $+150$ as a first value for the number
 of digits.
 In this way one can be sure that the result is good enough.
 A better way is to compute the polynomials twice, once with the default
 precision ($-1$), and once with a slightly better precision ($+10$).
 One can then compare the results which is printed after conversion to
 the internal 64-bit floating point datatype.
 The results should agree.\footnote{
 The ordering of the roots might change slightly.
 This is due to the fact that different orderings fullfil the same
 optimal ordering condition and it is just by chance which one of them
 is printed.
 However, the orderings should only differ by flips of root pairs with
 the same absolute value.}

 The output is written to files in the subdirectory specified by the
 last line of arguments.
 The file for the roots is named starting with {\em Coef}, and first in this
 file there are the coefficients of the polynomial and then 
 the real and imaginary parts of the roots.
 The file for the recurrence coefficients of this first polynomial 
 is named starting with {\em Cort}. If further polynomials are calculated
 there will be several files starting like that.
 Among these files the one for the first polynomial is the one not
 having a {\em -4} nor a {\em -5} near the end of the filename.
 In this file $(n+1)$ coefficients $d_\nu$ are given first, then
 $n$ values of $\beta_\mu$ and finally $(n-1)$ values of $\gamma_\mu$.

\subsection{The polynomials $P^{(2)}$ and $P^{(4)}$}\label{sec4.2}

 In TSMB applications $P^{(1)}_{n_1}$ is only a crude approximation to
 $f(x)=x^{-\alpha}$, i.~e.\ the polynomial order $n_1$ is quite small.
 This is corrected by a polynomial $P^{(2)}_{n_2}$ of higher degree
 $n_2$, which is defined such that $P^{(1)}P^{(2)}\approx f(x)$ to a
 better precision.
 This means that we set $\bar P = P^{(1)}$ and repeat the
 calculation as for $P^{(1)}$. 

 In a reweighting step the remaining deviation of $P^{(1)}P^{(2)}$ can
 be further corrected by a fourth polynomial $P^{(4)}_{n_4}$, defined
 similarly, which is computed by setting $\bar P = P^{(1)}P^{(2)}$.
 The polynomial degrees for $P^{(2)}_{n_2}$ and $P^{(4)}_{n_4}$ are
 given as the second and fourth argument in the first line of arguments.
 The weight function is set to the inverse itself.

 While the required number of digits for the first polynomial scales
 with a relatively small slope according to $(40+1.6 n_1)$, the
 corresponding slope is larger for the second and fourth polynomials
 because of the nested integrals introduced by setting $\bar P(x)$
 to more complicated functions.
 In our cases $(70+2.2 n_{2,4})$ gives a good estimate for the necessary
 number of digits.

 The recurrence coefficients $d_\nu$, $\beta_\mu$ and $\gamma_\mu$ for
 the second polynomial are appended to the output file containing the
 recurrence coefficients of the first polynomial.
 The ones for the fourth polynomial are written to a separate file with
 a name ending with {\em -4.c++}.

\subsection{The polynomial $P^{(3)}$}\label{sec4.3}

 For the TSMB algorithm, besides the before mentioned polynomials, one
 also needs an approximation $P^{(3)}_{n_3}$ of the inverse square root
 of $P^{(2)}_{n_2}$.
 This again is given by a polynomial labelled as the third one: 
\begin{equation} \label{eq33}
P^{(3)}_{n_3} \approx \left(P^{(2)}_{n_2}\right)^{-\frac{1}{2}}.
\end{equation}
 Since any error made in this polynomial approximation is uncorrectable
 in the TSMB updating algorithm, it is important that this approximation
 is good, especially at the lower end of the interval.
 To improve that part this polynomial is optimized for the interval
 $[\epsilon^\prime,\lambda]$, with $\epsilon^\prime=\frac{1}{10}\epsilon$.
 Furthermore one can adjust the weight function to $w(x)^2 = x^{-\omega_3}$.
 The polynomial degree $n_3$ for $P^{(3)}_{n_3}$ is read from the third
 argument in the second line of arguments, nominator and denominator
 of $\omega_3$ are given by the first two integer numbers in the 
 third line of arguments.

 The calculation of the third polynomial can be based on Neuberger's
 formula \cite{PRACTICAL}. The third value in the third line of arguments
 gives the number of terms taken in Neuberger's formula.
 For $K$ terms this polynomial is defined as:
\begin{equation}
 \label{eq34}
P^{(3)} \approx \frac{1}{K} \sum_{s=1}^K \frac{1}
{P^{(2)} \cos^2\frac{\pi}{2K}(s-\half) + \sin^2\frac{\pi}{2K}(s-\half)}
\ .
\end{equation}

 For large values of $K$ the computation using (\ref{eq34}) becomes
 slow.
 In this case one can use another method to obtain a good polynomial
 $P^{(3)}$ faster.
 This is implemented by a series of fifth polynomials $P^{(5)}_{n_5}$
 which are approxiamtions to a good $P^{(3)}$ with increasing
 precision. The same weight function as for the third polynomial is used.
 The members of this series are calculated iteratively.
 Each polynomial in the series may have a different polynomial degree,
 typically in a non-decreasing sequence.
 With this method it is also possible to further reduce the systematic
 error at some fixed polynomial order because the program uses a smaller
 lower bound for the approximation interval
 $\epsilon^{\prime\prime}=\frac{1}{100}\epsilon$.
 The polynomial degrees ($n_5$) of the $P^{(5)}_{n_5}$ sequence have to
 be given one after the other as the last arguments in the first line of
 arguments.

 The starting point for this iteration is some known approximation to
 the third polynomial $P^{(3)}_{(0)}$.
 From $P^{(3)}_{(0)}$ better approxiamtions can be obtained by the
 following Newton-type iteration:
\begin{equation}\label{eq35}
P^{(3)}_{(k+1)} = \frac{1}{2}\left(P^{(3)}_{(k)} + 
\frac{1}{P^{(3)}_{(k)}P^{(2)}}\right)
\qquad \quad k = 0, 1, 2, \ldots.
\end{equation}
 The implementation of this iteration calculates the term 
 $\frac{1}{P^{(3)}_{(k)}P^{(2)}}$ in the same way as the fourth
 polynomial is calculated, i.~e.\ there is a polynomial approximation
 there.

 After a few iterations, typically two or three, the quality of
 $P^{(3)}_{(k)}$ (also called ``fifth polynomial'') loses any dependence
 on the starting point of the iteration.
 Therefore it is also possible to replace the costly computation based
 on (\ref{eq34}) by some cruder approximation.
 One possibility is to take for $P^{(3)}_{(0)}$ a second polynomial
 ${P^{(2)}}^\prime$ for a first polynomial of ${P^{(1)}}^\prime$, which
 approximates $f^\prime(x)=x^{+\frac{\alpha}{2}}$ with the order $n_1$.
 Although this is not sufficient for using it elsewhere, it is very
 useful as a cheap starting point for the Newton iteration.
 This method is applied by the program if the specified number of
 terms for Neuberger's formula is negative.

 Present experience shows that using this crude estimate as a starting
 point is sufficient.
 One should then first create two or three fifth polynomials of lower
 order than the desired order for the final $P^{(5)}$.
 At the end one can increase the polynomial orders by larger steps.
 Usually a good approximation is achieved if the last polynomial
 order of $P^{(5)}_{n_5}$ is roughly $20\%$ larger than $n_2$.
 After the computation is done, the parameters of each fifth polynomial
 are stored in a seperate file.




\end{document}